\documentstyle[bo99,epsfig_latex209]{article}

\title{BeppoSAX  Observations of the   Galactic Center Region: 
Soft x--rays from the radio halo of SgrA East}

\author{L.~Sidoli $^{1,2}$, S.~Mereghetti $^1$, A.~Treves $^3$, 
L.~Chiappetti $^1$, G.L.~Israel $^4$ \& M.~Orlandini $^5$}                                           

\affil{1) Istituto di Fisica Cosmica `G. Occhialini", Milano, Italy  \\                  
2) Dipartimento di Fisica, Universit\`a di Milano, Milano, Italy \\                              
3) Universit\`a dell'Insubria, Como, Italy         \\                                        
4) Osservatorio Astronomico di Roma, Monteporzio Catone, Roma, Italy \\                            
5) ITeSRE, Bologna, Italy  }

\begin{document}

\def\deg {^\circ}
\def\mdot {\dot M}
\def\kms {$\sim$km$\sim$s$^{-1}$}
\def\gs {$\sim$g$\sim$s$^{-1}$}
\def\ergs {$\sim$erg$\sim$s$^{-1}$}
\def\cmtre {$\sim$cm$^{-3}$}\def\nupa{\vfill\eject\noindent}
% definitions ----------------------------------------------------

\maketitle

\begin{abstract}
The   $Beppo$--SAX satellite performed a survey of 
the   Galactic Center Region 
in the 1-10 keV energy band with its  Narrow Field Instruments.
Several bright X--ray sources containing neutron stars and black holes  have
been observed and studied, including the possible counterpart of 
SgrA*.   
Here we report  the results on the diffuse emission coming 
from the SgrA Complex.
 
The emission from within  8$'$ from SgrA*  has 
a  double--temperature  thermal spectrum (kT$_{1}\sim$ 0.6 keV  
and  kT$_{2}\sim$8~keV) and  an  energy--dependent morphology:
the hard emission (5--10 keV)  is   elongated along the 
galactic plane, while the  soft one (2--5 keV) shows 
a  triangular shape, very similar to the  radio   halo of  SgrA~East. 
This  spatial correlation and the physical parameters
of the   lower temperature  
component  support the interpretation 
of the   radio halo of the Sgr~A~East shell as a SNR.

\end{abstract}

\section{Introduction}

The   BeppoSAX satellite performed a survey of 
the   Galactic Center Region 
in the 1--10 keV energy band with its  Narrow Field Instruments during 1997--1998. 

A source positionally coincident with the Galactic Center (hereafter GC) 
was observed, together with strong
diffuse emission and several point--like sources with luminosity 
L$_{X}\sim10^{36}$~erg~s$^{-1}$. 
The results on these  sources, most of which are likely
low mass X--ray binaries containing neutron
stars and black holes, both with transient and persistent emission,
are reported in detail by  Sidoli et al. (1999).
The spectral results for these
 sources are summarized in Fig.~1, where the photon index 
of the fits with a power law are plotted versus the hydrogen column density. 
An upper limit of 
L$_{X}\sim10^{35}$~erg~s$^{-1}$ has also been placed  to the 2--10 keV luminosity
from the X--ray counterpart of SgrA*  (see Sidoli et al. 1999 for details),
confirming the underluminosity of this presumed supermassive   black hole at
high energies.  

Intense diffuse X--ray emission  is also present in the GC region, the
nature of which is still poorly known.  Here we present the 
BeppoSAX results on  the diffuse emission from  the  SgrA complex
(Sidoli \& Mereghetti 1999).

\begin{figure}
\vskip -5truecm
\centerline{\psfig{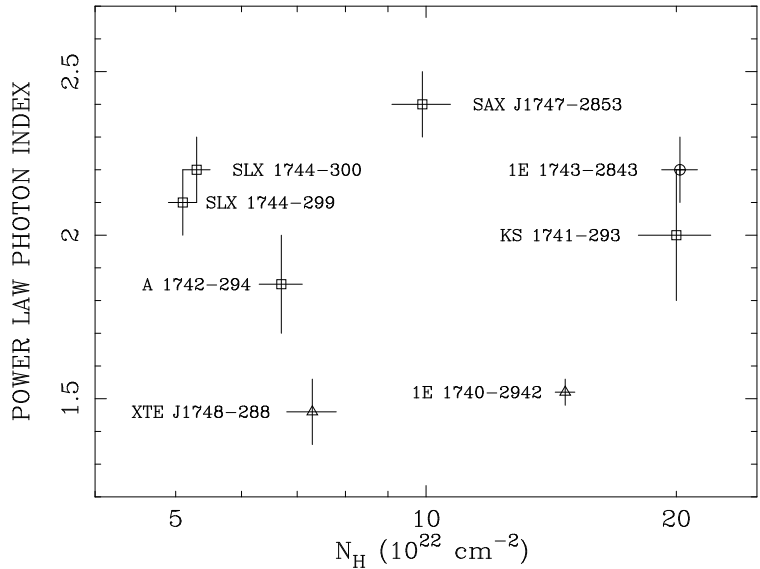}}
\vskip -1truecm
\caption[]{Spectral slope of the brightest sources observed in the GC region. 
Triangles mark the black hole candidates, the squares indicate 
LMXBs with neutron stars, while the circle
 marks a source the nature of which is still unknown.}
\end{figure}

\section{The Diffuse Emission from the SgrA Complex}

The Sgr~A Region has been imaged with the MECS instruments (1.3--10 keV) 
with a spatial resolution of $\sim1'$ (about 2.5 pc at the GC distance)
in August 1997   (99.5 ksec net exposure time).

In order to study the spectral properties of the 
diffuse emission coming from the SgrA Complex, 
we extracted the  MECS   counts  from  four concentric annular regions 
($0'-2'$,$2'-4'$,$4'-6'$,$6'-8'$) around SgrA*.

Several emission lines are present in all the spectra, 
with the K-lines from iron (E$\sim6.7$~keV) 
and sulfur (E$\sim2.4$~keV) particularly bright.
The fit with a single temperature hot plasma  model showed a 
nearly constant temperature ($\sim7-8$ keV)   at radii larger than 
2$'$, while in the innermost circle   a softer
spectrum (kT$\sim4$ keV) was found.
This is probably due to the contribution   
 from one (or more)   point source(s) located close to SgrA* 
 (Predehl \& Tr\"umper 1994, Maeda et al. 1996, 
Sidoli et al. 1999), that cannot
be spatially resolved in our data.
Since the temperature profile does not
show   spectral variations in the region 
from $2'$ to $8'$, we studied the overall spectrum from  this entire corona.
  A single temperature   plasma (MEKAL model) is not adequate to describe the spectrum,
leaving positive residuals at low energy and especially around  6.4 keV.
This can be due to the presence of fluorescent emission from 
neutral or weakly ionized iron 
in  the nearby molecular clouds (Koyama et al. 1996).
Thus we added a lower temperature plasma component plus a gaussian line at 6.4 keV.
The resulting best fit is a double--temperature   plasma, with  
kT$_{1}\sim$0.6~keV and kT$_{2}\sim$8~keV 
 (N$_{H}$=8$\times10^{22}$~cm$^{-2}$) and a gaussian line at 6.4 keV with an equivalent
 width of   $\sim$120 eV (see Fig.~2).
The total flux corrected for the absorption is 
F$_{X}\sim1.7\times10^{-10}$~erg~cm$^{-2}$~s$^{-1}$ (2--10 keV), which
translates into a luminosity of  L$_{X}\sim1.4\times10^{36}$~erg~s$^{-1}$. 
About one third of the flux is contributed by the soft component.

%--------------------------  figure 2
\begin{figure}
\vskip -3.5truecm
\centerline{\psfig{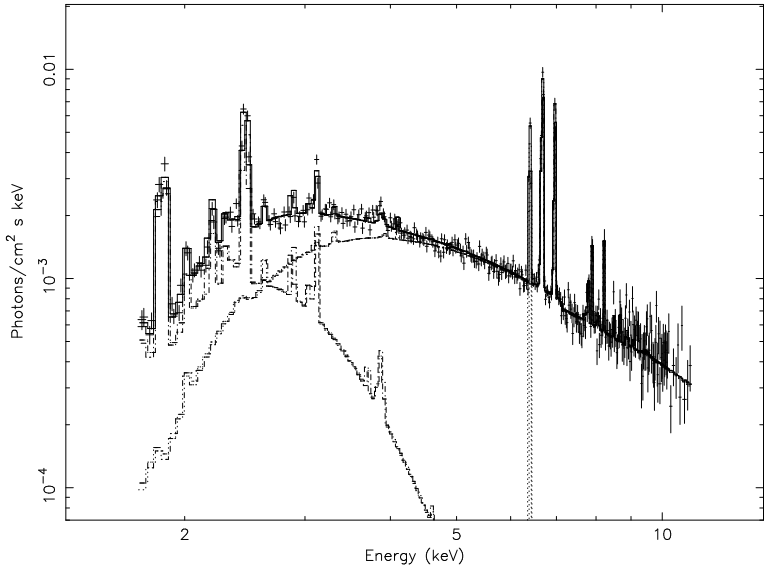}}
\vskip -2truecm
\caption[]{Best fit to the MECS spectrum (2$'-8'$ corona) from the Sgr~A Complex.} 
\end{figure}
%---------------------------------

\subsection{Morphology of the Diffuse Emission}

%--------------------------  figure 3
\begin{figure}
\vskip -2truecm
\centerline{\psfig{file=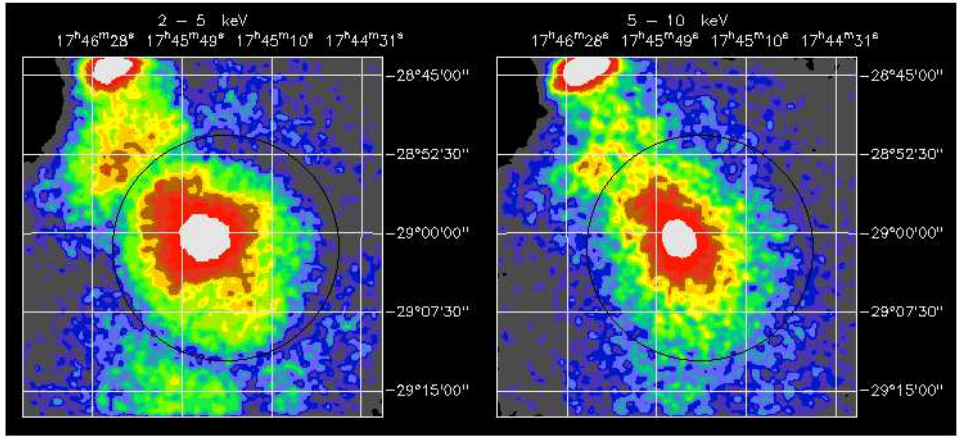,width=13.0cm}}
\caption[]{Spatial distribution of the diffuse emission from the Sgr~A Complex below (left panel)
and above (right panell) 5 keV. A smoothing with gaussian with FWHM=1$'$ has been applied.
The black circle marks the position of the circular strongback structure of the MECS instrument 
(about 20$'$ diameter).} 
\end{figure}
%---------------------------------

The spatial distribution of the diffuse emission  
was studied extracting two images in different energy ranges, 
below and above 5 keV (see Fig.~3). 
Both   emissions are peaked at the GC position, but they
have  significantly different spatial distributions: the soft emission
(2--5 keV) displays  a triangular shape,  while the hard one (5--10 keV) 
is elliptical and elongated
in the direction of the galactic plane.
While the hard emission can be simply part of the diffuse emission permeating
the inner 60$^{\circ}$ of the galactic disk  (e.g. Kaneda  et al. 1997), the soft X--rays
are spatially  correlated  with a   structure observed in the radio band,
known as the
Sgr~A East triangular halo (Pedlar et al. 1989).
This is  an extended  non--thermal  
structure (probably a SNR) which  surrounds in projection SgrA*.
Since also our spectral data are well described  by a   two--temperature 
thermal model, it is tempting
to give an interpretation in terms of two plasma components with 
 different temperatures and   spatial distributions. 
From  the  spectral analysis of
the lower temperature plasma   (kT$_{1}\sim0.6$~keV),
we derive a luminosity $\sim4.5\times10^{35}$~ergs~s$^{-1}$  (2--10 keV), 
an electron density n$_{e}\sim3$~cm$^{-3}$, a total
mass $M_{g}\sim250~M_{\odot}$ and  an  average thermal 
pressure $P\sim3\times10^{-9}$~ergs~cm$^{-3}$, consistent 
with the pressure $P_{Sedov}\sim4\times10^{-9}$~erg~cm$^{-3}$
derived  for a SNR   in a Sedov phase.  

\section{Conclusions}

The BeppoSAX observation of the SgrA complex revealed  the presence 
of at least 
two distinct components: a soft component  with kT$\sim$0.5--1 keV 
spatially correlated with the SgrA~East $7'$ triangular halo, 
and a hard one with  kT$\sim$7--9 keV elongated  along the 
Galactic Plane and  possibly associated with the harder component of the
Galactic Ridge  emission.

The soft component, which accounts for about one third of
the 2-10 keV diffuse luminosity from the SgrA complex, 
can be well explained as thermal emission from the SNR
responsible for the radio halo of the SgrA East shell.

\end{document}